\documentclass{aastex}

\usepackage{graphicx}

\usepackage{emulateapj5}
\usepackage{times}

\usepackage{rotating}

\makeatletter

\newenvironment{inlinefigure}{%
\def\@captype{figure}%
\noindent\begin{minipage}{0.999\linewidth}\begin{center}}
{\end{center}\end{minipage}\smallskip}
\makeatother

\usepackage{mathptmx}

\def\Mo     {{\rm M}_{\odot}}

\begin{document}


\title{The Universal Gas Mass Fraction in Clusters of Galaxies}


\author{Laurence P. David, Christine Jones \& William Forman}
\affil{Harvard-Smithsonian Center for Astrophysics, 60 Garden St., Cambridge, MA 02138}
\email{ldavid@head.cfa.harvard.edu}


\begin{abstract}
We obtained a deep 150~ksec {\it Chandra} observation of the
optically selected cluster of galaxies, RCS 2318+0034, to investigate
the gas mass fraction in this system. Combining our deep {\it Chandra}
observation with an archival 50~ksec observation, we derive
gas mass fractions of $f_{gas}=0.06 \pm .02$ and $0.10 \pm .02$
within $r_{2500}$ and $r_{500}$, respectively.
The gas mass fraction in RCS 2318+0034 within $r_{500}$ is
typical of X-ray selected clusters.  Further study shows that
the large scale properties of RCS 2318+0034, including
the relations between gas mass, X-ray luminosity and
gas temperature are also consistent with the observed correlations
of X-ray selected clusters.  However, the gas mass fraction
within $r_{2500}$ is less than most X-ray selected clusters,
as previously reported.  The deep {\it Chandra} image of RCS 2318+0034
shows that this system is presently undergoing a major
merger which may have an impact on the inferred
gas mass fraction within $r_{2500}$.

\end{abstract}


%% subject header.  Objects should be in the appropriate "individual"
%% headers (e.g. quasars: individual, stars: individual, etc.) with the
%% additional provision that the total number of headers, including each
%% individual object, not exceed six.  The \objectname{} macro, and its
%% alias \object{}, is used to mark each object.  The macro takes the object
%% name as its primary argument.  This name will appear in the paper
%% and serve as the link's anchor in the electronic edition if the name
%% is recognized by the data centers.  The macro also takes an optional
%% argument in parentheses in cases where the data center identification
%% differs from what is to be printed in the paper.


\section{Introduction}

The current matter in rich clusters of galaxies was accumulated from
regions that would presently span approximately 10~Mpc and should
comprise a fair sample of the Universe.  It has been known for
some time that the dominant component of baryonic matter in rich
clusters is the hot X-ray emitting gas
(David et al. 1990; White \& Fabian 1995; David et al. 1995;
Jones \& Forman 1999; Ettori et al. 2003; Allen et al. 2004; Vikhlinin et al. 2006;
Allen et al. 2008; Sun et al. 2009; Vikhlinin et al. 2009).  White et al. (1993)
showed that the observed high gas mass fraction in clusters has
significant cosmological consequences regarding the total mass
density of the universe.  The segregation of baryons between stars
and hot gas depends on both the radius within a group or cluster
and the total mass of the system.  The ratio of gas mass to
stellar mass increases from approximately unity in the cores of
groups up to more than than a factor of five at large radii in the
richest and hottest clusters (David et al. 1995, Gonzales et al. 2007).
In addition, since the gas is more extended than the dark matter,
the gas mass fraction increases with radius within
groups and clusters
(David et al. 1995, White \& Fabian 1995, Allen et al. 2004,
Vikhlinin et al. 2006).  The greater extent of the gas, relative to
the stellar and dark matter, probably arises from some form of
non-gravitational heating (e.g., mechanical AGN heating  and
supernova driven galactic winds) which breaks the self-similarity
between groups and clusters
(Ponman et al. 1999, Arnaud \& Evrard 1999, Lloyd-Davies et al. 2000,
Bialek et al. 2001, Neumann \& Arnaud 2001, Finoguenov et al. 2002,
Voit et al. 2005, Kravtsov et al. 2005, Dav\'{e} et al. 2008,
Bode et al. 2009).

Most studies of nearby clusters find that the gas mass fraction
approaches a nearly constant value of $f_{gas} \approx 13\%$ at
large radii in the hottest and most massive clusters (David et al. 1995,
White \& Fabian 1995, Allen et al. 2004, Vikhlinin et al. 2006,
Sun et al. 2009). This is slightly less than the WMAP
value of 17\%.  Based on these findings, there have been several
cosmological studies whose results are based on the assumption that
the gas mass fraction in rich clusters is independent of redshift
(Ettori et al. 2003, Allen et al. 2004, Ferramacho \& Blanchard 2007,
Ettori et al. 2009, Juett et al. 2010).
All of these studies have found results consistent
with the concordant $\Lambda$CDM cosmological model.
However, a recent analysis of {\it Chandra} observations of 13 optically
selected, moderate redshift clusters culled from the Red-Sequence Cluster
Survey (RCS; Gladders \& Yee 2005) by
Hicks et al. (2008), reports a significant scatter in the gas mass
fraction of clusters with average gas mass fractions of
$f_{gas}$ = 4.8\%  at $r_{2500}$ and 6.8\% at $r_{500}$
(where $r_{\Delta}$ is the radius within which the average mass
density of the cluster is $\Delta$ times the critical density of
the universe at the redshift of the cluster).
If this result is more representative of clusters compared to
previous studies of X-ray selected clusters, then it would
make precision cosmological studies based on the assumption of a
universal gas mass fraction in clusters problematic.
The cluster with the lowest gas mass fraction in the Hicks
et al. (2008) sample is RCS 2318+0034 with reported
values of $f_{gas}$=4.0\% and 1.5\% at $r_{2500}$
and $r_{500}$, respectively.  Due to the cosmological consequences
of this result, we obtained an additional 150~ksec {\it Chandra}
observation of RCS 2318+0034 in AO-11 to better constrain
the gas mass fraction in this system.

This paper is organized as follows.  Section 2 contain a
detailed discussion about the {\it Chandra} data analysis. Section
3 presents evidence for a recent merger in RCS 2318+0034 and
$\S$4 examines the X-ray surface brightness distribution.
All spectral analysis of the {\it Chandra} data is presented
in $\S$5.  Section 6 gives the gas mass,
total gravitating mass and gas mass fraction at $r_{2500}$ and $r_{500}$ and
a comparison of our work with previous estimates is given
in $\S$7.  Our main results and their cosmological implications
are summarized in $\S$8.

\section{Data Analysis}

RCS 2318+0034 was observed with the {\it Chandra} ACIS-S detector on May 3,
2005 (ObsID 4938) and with the ACIS-I detector on Aug. 7, 2009
(ObsID 11718). Both observations were taken in very faint (VF) telemetry mode.
All data were re-processed with CIAO 4.3 and CALDB 4.4.6.
We used {\it acis\_process\_events} to apply the latest CTI and gain corrections
and filtered the VF mode data to reduce the charged particle background.
During the ACIS-S observation, the I2, I3, S1, S2 and S3
chips were turned on and the cluster was positioned on 

\begin{figure*}
\includegraphics[width=1.00\textwidth,bb=43 299 520 474,clip]{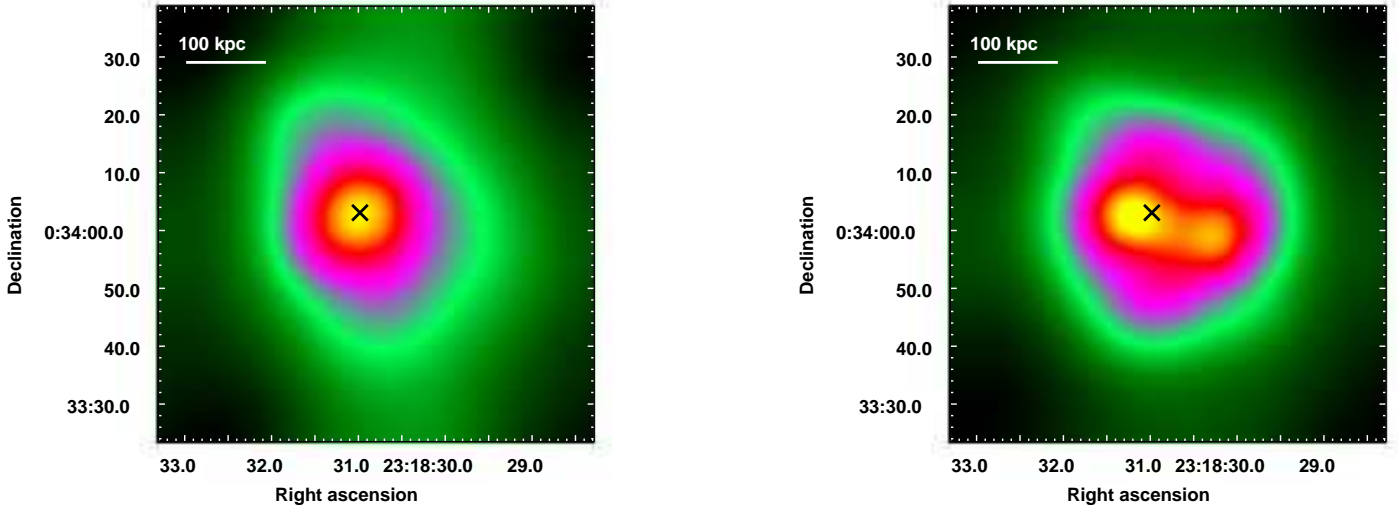}
\caption{left: hard band (2.0-6.0~kev) adaptively smoothed image, and right: 
soft band (0.5-2.0~keV) adaptively smoothed image of RCS 2318+0034.
The cross marks the location of the peak surface brightness of the 
main cluster in the hard band image.}
\end{figure*}

\noindent
the S3 chip.
During the ACIS-I observation, the I0, I1, I2,
I3 and S3 chips were turned on and the cluster was positioned on the I3 chip.
Light curves were generated in the 2.5-7.0~keV and 9.0-12.0~keV energy bands
for the BI chip S1 during the ACIS-S observation and for the BI chip S3
during the ACIS-I observation. All time intervals with background
rates in these energy bands exceeding 20\% of the quiescent background
rate were excised from further analysis.  The resulting cleaned
exposure times for the ACIS-S and ACIS-I observations
are 50,155sec and  143,275sec. The appropriate ACIS blank
field background files (Epoch D for the ACIS-S observation and Epoch E for the
ACIS-I observation) were extracted from the CALDB and
reprojected onto the sky to match the two RCS 2318+0034 observations.
The same VF background screening was applied to the background data sets
by only including events with "status=0". Finally, the exposure times
for each chip in the background data sets were adjusted to produce the
same 9-12~keV count rates as in the two RCS 2318+0034 observations.

Molecular contamination has been building up on the cold ACIS
filters since launch.  This only affects the ACIS QE below 2~keV.
Since the ACIS-I observation of RCS 2318+0034 was
taken more recently than the observations used to compile the
background files, the throughput of ACIS at low energies in
the RCS 2318+0034 observation is less than it was during the
background observations.  To correct for this affect, we
extracted spectra from a source free region on I3 beyond
the virial radius of the cluster in the ACIS-I
observation of RCS 2318+0034 and the I3 blank field background file.
We found that the count rate in the 0.3-6.0~keV energy band
(i.e., the energy band we use to generate surface brightness profiles)
during the RCS 2318+0034 observation is still 4.6\% less than
the count rate in the I3 blank field data set even after adjusting
the exposure time in the background data set to produce
the same 9-12 keV count rate.  We therefore
reduced the count rate in the I3 background data set
by 4.6\% before subtracting the background.
We repeated this procedure
for all 5 chips (all 4 ACIS-I chips from the ACIS-I
observation and the S3 chip from the ACIS-S observation)
used to generate background-subtracted surface brightness
profiles for RCS 2318+0034.  For all spectral analysis, we used
the CIAO task {\it specextract} which extracts source and
background spectra and computes photon-weighted effective
area and response files.

Due to the lack of measured galaxy redshifts for RCS 2318+0034,
we adopt a redshift of z=0.78 (which is the same as that used
in Hicks et al. 2008) derived from the energy centroid
of the H-like and He-like Fe-K$\alpha$ emission lines in the ACIS spectrum.
Assuming the standard $\Lambda$CDM cosmology with
$\rm H_0$=70~km~s$^{-1}$~Mpc$^{-1}$, $\Omega_{M}$=0.3
and $\Omega_{\Lambda}$=0.7, gives a luminosity distance of
$D_L=4,870$~Mpc to RCS 2318+0034 and $1^{\prime\prime} = 7.45$~kpc
in the rest frame of the cluster.

\section{X-Ray Morphology}

Adaptively smoothed ACIS images of the co-added RCS 2318+0034
observations in a hard
(2.0-6.0~keV) and soft (0.5-2.0~keV) energy band
are shown in Figure 1. In the hard band image,
RCS 2318+0034 appears to be a fairly relaxed hot cluster, while the soft
band image reveals that RCS 2318+0034 is undergoing a merger with
a smaller and cooler system that is presently located about 100~kpc
to the west of the dominant cluster.  The "x" in both figures
indicates the location of the peak in the hard band surface brightness of the
main cluster.  The soft energy band image shows that the cool gas in the
center of the main cluster has been displaced toward the
east during the merger, possibly due to merger-induced sloshing
(Ascasibar \& Markevitch 2006).

Since there are insufficient counts in the ACIS data to generate a
temperature map, we generated a hardness ratio image by
adaptively smoothing the raw 0.5-2.0~keV and 2.0-6.0~keV
images with the same kernel and then divided the two images.
The resulting hardness ratio image is shown in Figure 2 along with
surface brightness contours derived from
the soft band image shown in Figure 1.  The hardness ratio image
shows that there is an elongated region of softer X-ray emission
associated with the merging cooler system.  The coolest emission from
the main cluster is slightly displaced toward the east from
the cluster core.  The hardest X-ray emission arises from the
region between the two merging cores.  The spectral analysis presented
below shows that this hard X-ray emission probably originates
from a merger induced shock.

\begin{inlinefigure}
\center{\includegraphics*[width=1.00\linewidth,bb=67 140 517 624,clip]{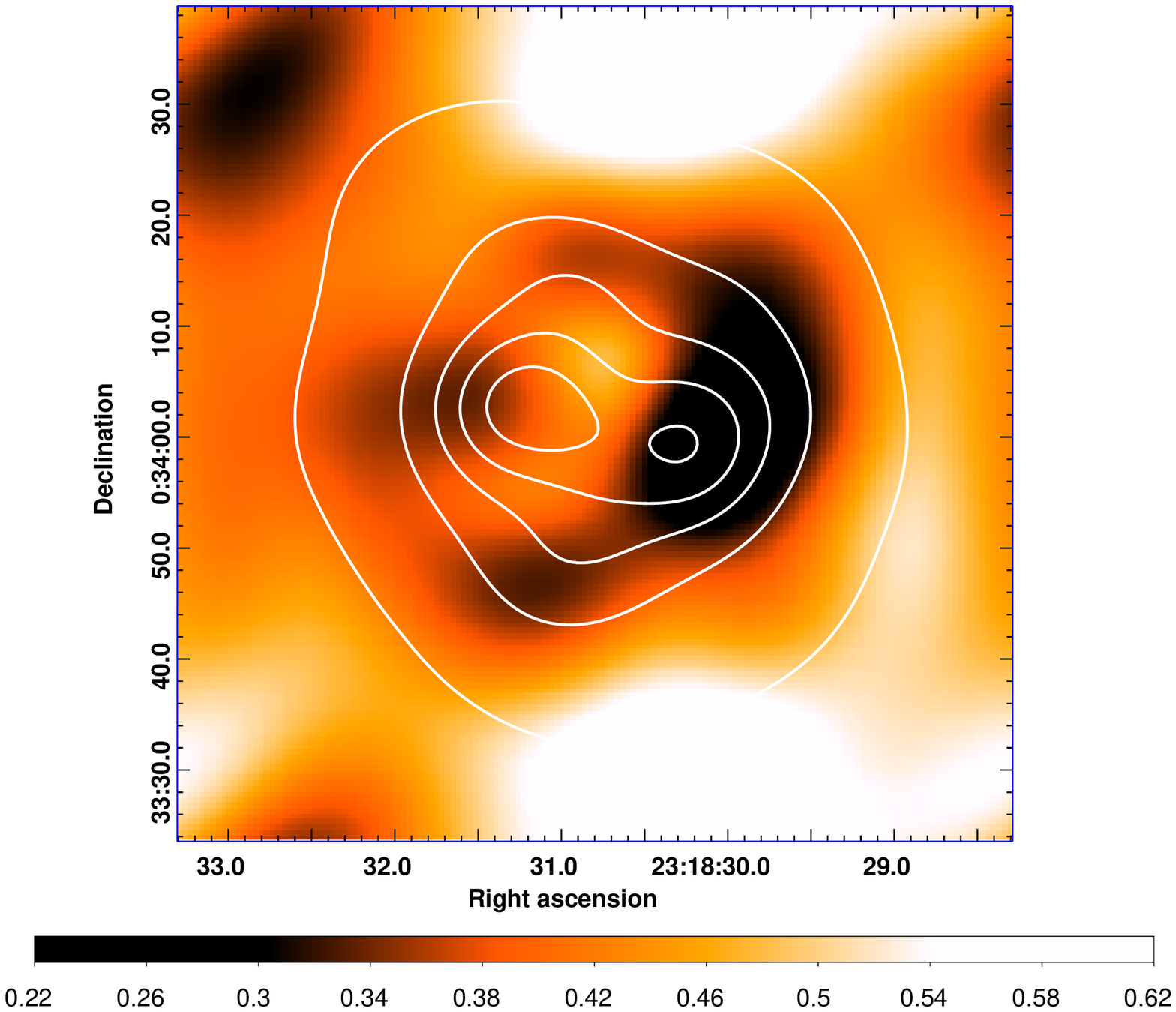}}
\caption{Hardness ratio image along with surface brightness contours derived from the
soft band image shown in Fig. 1.}
\end{inlinefigure}

\section{Surface Brightness Profile}

Due to the on-going merger in RCS 2318+0034, we extracted surface brightness
profiles using both full annuli and annuli excluding a 120$^{\circ}$ pie
slice toward the west.  All surface brightness profiles were
computed using the location
of the cross in Figure 1 as the origin. Data
from the ACIS-S and ACIS-I observations were co-added and
the normalized blank sky background images were used for background
subtraction.  To determine the appropriate weighting for the instrument and
exposure maps, we extracted a spectrum from the central 500~kpc
of RCS 2318+0034 and fit the spectrum to a single temperature
{\it apec} model with the absorption fixed at the galactic
value ($N_{gal}=4.04 \times 10^{20}$~cm$^{-2}$) and obtained
kT=7.5 (6.4-8.6)~keV, an abundance of $Z=0.12 (<0.33)$
(based on the abundance table of Grevesse \& Sauval 1998)
and an emission-measure of $\epsilon = (3.21 \pm 0.16) \times 10^{-4}$cm$^{-5}$.
All error bars are given at the 1$\sigma$ confidence level.
Instrument and exposure maps weighted by the thermal emission from
a 7.5~keV plasma were then generated for all 5 CCDs used to
calculate the surface brightness profiles.
Figure 3 shows the resulting background-subtracted and exposure-corrected
0.5-6.0~keV surface brightness profile excluding the emission
from a 120$^{\circ}$ pie slice toward the west.  We refer to
this as our standard surface brightness profile.

The X-ray surface brightness profile of clusters is commonly
fit with a $\beta$ model, given by:

\begin{equation}
\Sigma(r) = \Sigma_0 \left( 1 + \left( {{r} \over {r_c}} \right)^2 \right)^{-3 \beta +1/2}
\end{equation}

\noindent
The results of fitting a $\beta$ model to the background-subtracted
and exposure-corrected surface brightness profiles
using full annuli are shown in Table 1.
This Table shows that as larger radii are included in the
fitting process, larger values of $\beta$ and core radii,
$r_c$, are obtained.  The results of fitting the surface 

\begin{inlinefigure}
\center{\includegraphics*[width=1.00\linewidth,bb=20 145 569 695,clip]{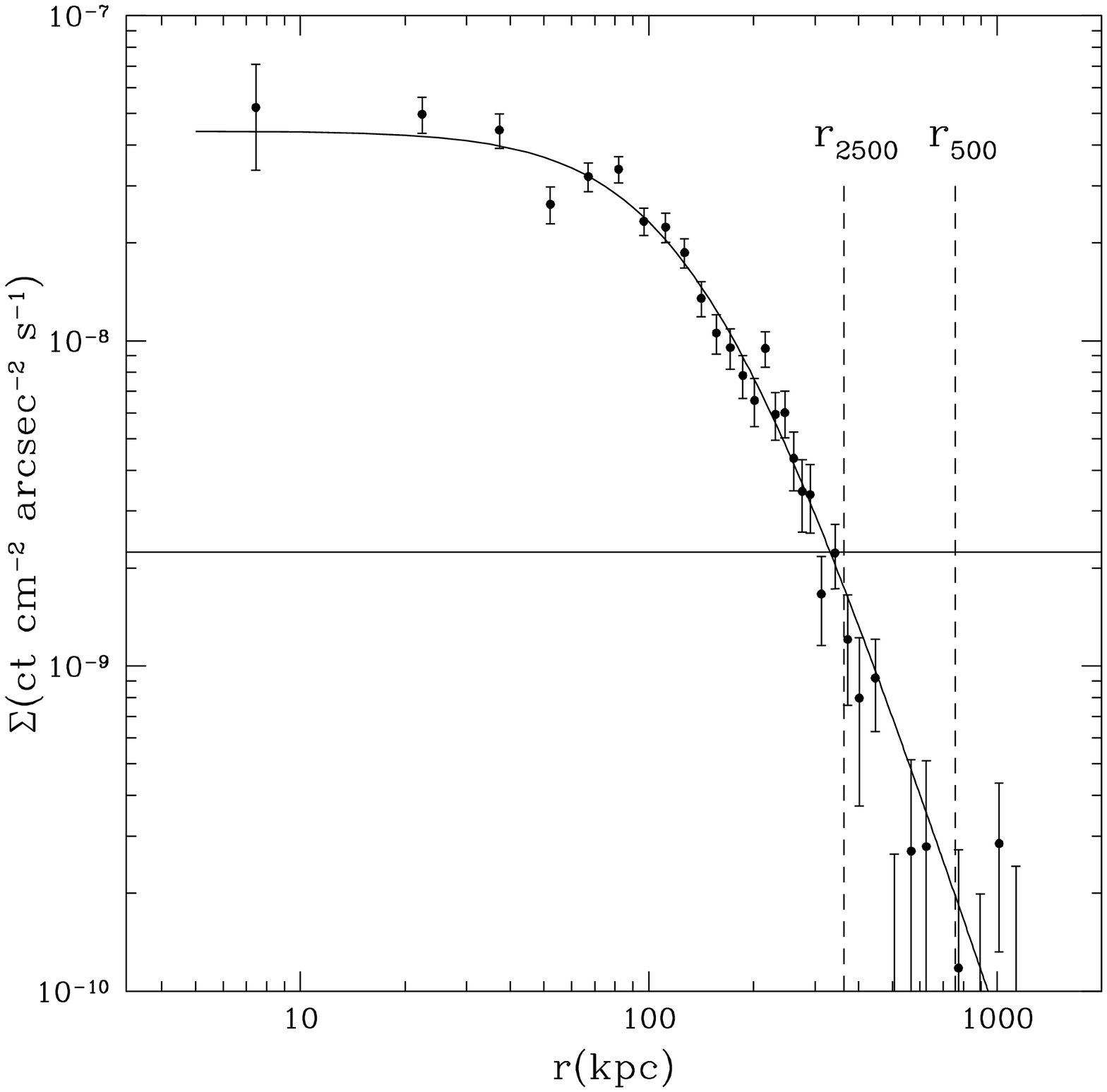}}
\caption{Background-subtracted and exposure-corrected 0.3-6.0~keV surface
brightness profile of RCS 2318+0034 excluding the emission from a 120$^{\circ}$ pie
slice toward the west along with our standard $\beta$ model (solid line).
also shown are the ACIS-I background (horizontal
solid line) and $r_{2500}$ and $r_{500}$ (vertical dashed lines).}
\end{inlinefigure}

\noindent
brightness profiles that
exclude the emission from a 120$^{\circ}$ pie slice are
shown in Table 2. This table shows that smaller
values of $\beta$ and $r_c$ are obtained when the emission
from the merging system toward the west is excluded,
however, there is still a trend of increasing $\beta$
and $r_c$ when larger regions are included in the fitting
process.  The main difficulty in fitting the surface brightness
profile is that the X-ray surface brightness of
RCS 2318+0034 drops below the I3 background around
350~kpc, which is only about 2 core radii from the center of the
cluster (see Figure 3).  It is therefore difficult to obtain
a tight constraint on the asymptotic slope of the X-ray surface
brightness.

To explore the systematic uncertainties in the derived parameters of
the $\beta$ model, we generated a surface brightness profile
excluding the emission from a 120$^{\circ}$ pie slice,
using an exposure map weighted by the thermal emission from a 4~keV
thermal plasma, since the gas temperature in clusters typically decreases
by a factor of 2 at large radii (Vikhlinin et al. 2006). Table 2 shows
that the derived values for the $\beta$ model are consistent
with the values obtained using an exposure map weighted by the thermal emission
from a 7.5~keV plasma.  We also varied the background level
by $\pm$5\% to determine the sensitivity of the derived
parameters on the estimated background.
This exercise shows that if the surface brightness profile is
only fit within the central 350~kpc, then the results are fairly
insensitive to the adopted background.
However, if the emission from larger radii is included in the analysis,
the results are quite sensitive to the exact background level used
in the analysis.

\section{Spectral Analysis}

To further investigate the on-going merger in the center of
RCS 2318+0034, we extracted spectra from the two cores and
the region between the two cores. The results of fitting
a single temperature {\it apec} model with the hydrogen column
density fixed at the galactic value are shown in Table 3.
While the uncertainties on the gas temperatures are fairly large,
due to the 

\begin{table*}[t]
\begin{center}
\caption{Surface Brightness Profile - Full Annuli}
\begin{tabular}{lccc}
\hline
Fitted Region & $\beta$ & $\rm{r_c}$ & $\Sigma_0$ \\
(kpc) & &  (kpc) & (ct~cm$^{-2}$~arcsec$^{-2}$~sec$^{-1}$) \\
\hline\hline
0-250 & $0.64 \pm .11$ & $136 \pm 28$ & $(4.53 \pm 0.30) \times 10^{-8}$ \\
0-350 & $0.74 \pm .09$ & $159 \pm 23$ & $(4.42 \pm 0.26) \times 10^{-8}$ \\
0-450 & $0.85 \pm .09$ & $183 \pm 21$ & $(4.29 \pm 0.23) \times 10^{-8}$ \\
\hline
\end{tabular}
\end{center}
\noindent
Notes:  Results of fitting a $\beta$ model within different
radii (Fitted Region) to the surface brightness profile extracted from
within full annuli.  The
nominal background rates and instrument maps
weighted by the emission from a 7.9~keV thermal plasma were used in
the analysis.
\end{table*}

\begin{table*}[t]
\begin{center}
\caption{Surface Brightness Profile - Excluding a 120$^{\circ}$ Pie Slice}
\begin{tabular}{lccc}
\hline
Fitted Region & $\beta$ & $r_c$ & $\Sigma_0$ \\
(kpc) & &  (kpc) & (ct~cm$^{-2}$~arcsec$^{-2}$~sec$^{-1}$) \\
\hline\hline
0-250 & $0.58 \pm .09$ & $112 \pm 27$ & $(4.63 \pm 0.41) \times 10^{-8}$ \\
0-350$^{\dag}$ & $0.70 \pm .09$ & $142 \pm 23$ & $(4.42 \pm 0.33) \times 10^{-8}$ \\
0-450 & $0.75 \pm .08$ & $152 \pm 21$ & $(4.36 \pm 0.23) \times 10^{-8}$ \\
\hline
0-250$^a$ & $0.58 \pm .09$ & $112 \pm 26$ & $(5.39 \pm 0.47) \times 10^{-8}$ \\
0-350$^a$ & $0.70 \pm .09$ & $141 \pm 23$ & $(5.07 \pm 0.38) \times 10^{-8}$ \\
0-450$^a$ & $0.75 \pm .08$ & $152 \pm 21$ & $(4.99 \pm 0.36) \times 10^{-8}$ \\
\hline
0-250$^b$ & $0.59 \pm .09$ & $114 \pm 27$ & $(4.60 \pm 0.40) \times 10^{-8}$ \\
0-350$^b$ & $0.73 \pm .10$ & $146 \pm 25$ & $(4.39 \pm 0.33) \times 10^{-8}$ \\
0-450$^b$ & $0.80 \pm .09$ & $161 \pm 23$ & $(4.30 \pm 0.31) \times 10^{-8}$ \\
\hline
0-250$^c$ & $0.57 \pm .09$ & $110 \pm 26$ & $(4.65 \pm 0.41) \times 10^{-8}$ \\
0-350$^c$ & $0.68 \pm .08$ & $137 \pm 22$ & $(4.46 \pm 0.34) \times 10^{-8}$ \\
0-450$^c$ & $0.71 \pm .07$ & $144 \pm 19$ & $(4.41 \pm 0.32) \times 10^{-8}$ \\
\hline
\end{tabular}
\end{center}
\noindent
Notes:  Results of fitting a $\beta$ model within different
radii (Fitted Region) to surface brightness profiles extracted from annuli
that exclude the emission from a 120$^{\circ}$ pie slice toward the west.
The top 3 rows give the results using instrument maps weighted by the emission
from a 7.5~keV thermal plasma. $^a$ Results using instrument maps weighted by
the emission from a 3~keV thermal plasma.  $^b$ Results obtained by increasing
the background rates by 5\%. $^c$ Results obtained by decreasing the background
rates by 5\%. $^{\dag}$ Our standard $\beta$ model.
\end{table*}

\begin{table*}[t]
\begin{center}
\caption{Spectral Analysis of Selected Regions}
\begin{tabular}{lc}
\hline
Region & kT \\
&  (keV)  \\
\hline\hline
Western Core & 3.24 (2.70-4.19) \\
Shock & 11.5 (7.27-21.6) \\
Eastern Core & 7.18 (4.36-14.7) \\
\hline
\end{tabular}
\end{center}
\noindent
Notes: Best-fit temperatures and $1\sigma$ uncertainties.
\end{table*}

\noindent
limited photon statistics, we find that the
western core is cooler than the eastern core, as inferred from
the hardness ratio map.  There is also a region of hotter gas
between the two cores.  This is consistent with a merger scenario in which
the hotter gas was produced by a merger shock.  The large
uncertainties in the temperature prevent a more detailed analysis
of the merger dynamics.

A temperature profile was generated by extracting the emission
from within 3 concentric annuli with approximately 1,000 net
counts per spectrum.  The spectra were fit
to an absorbed single temperature {\it apec} model with the hydrogen column
density fixed at the galactic value.  The resulting
temperature profile shows some evidence for a
radial decrease in the gas temperature (see Figure 4).
Also shown in Figure 4 is the broken power-law temperature profile
used to estimate the total gravitating mass in RCS 2318+0034
and the resulting values for $r_{2500}$ and $r_{500}$ (see \S 6).
In addition, we extracted a spectrum from the innermost region
excluding the emission within a 120$^{\circ}$ pie slice toward the
west (which excises the emission from the merging subcluster), but the
resulting gas temperature is consistent with the inner temperature
shown in Figure. 4. Since the emission
from RCS 2318+0034 drops below the X-ray 

\begin{inlinefigure}
\center{\includegraphics*[width=1.00\linewidth,bb=20 145 584 695,clip]{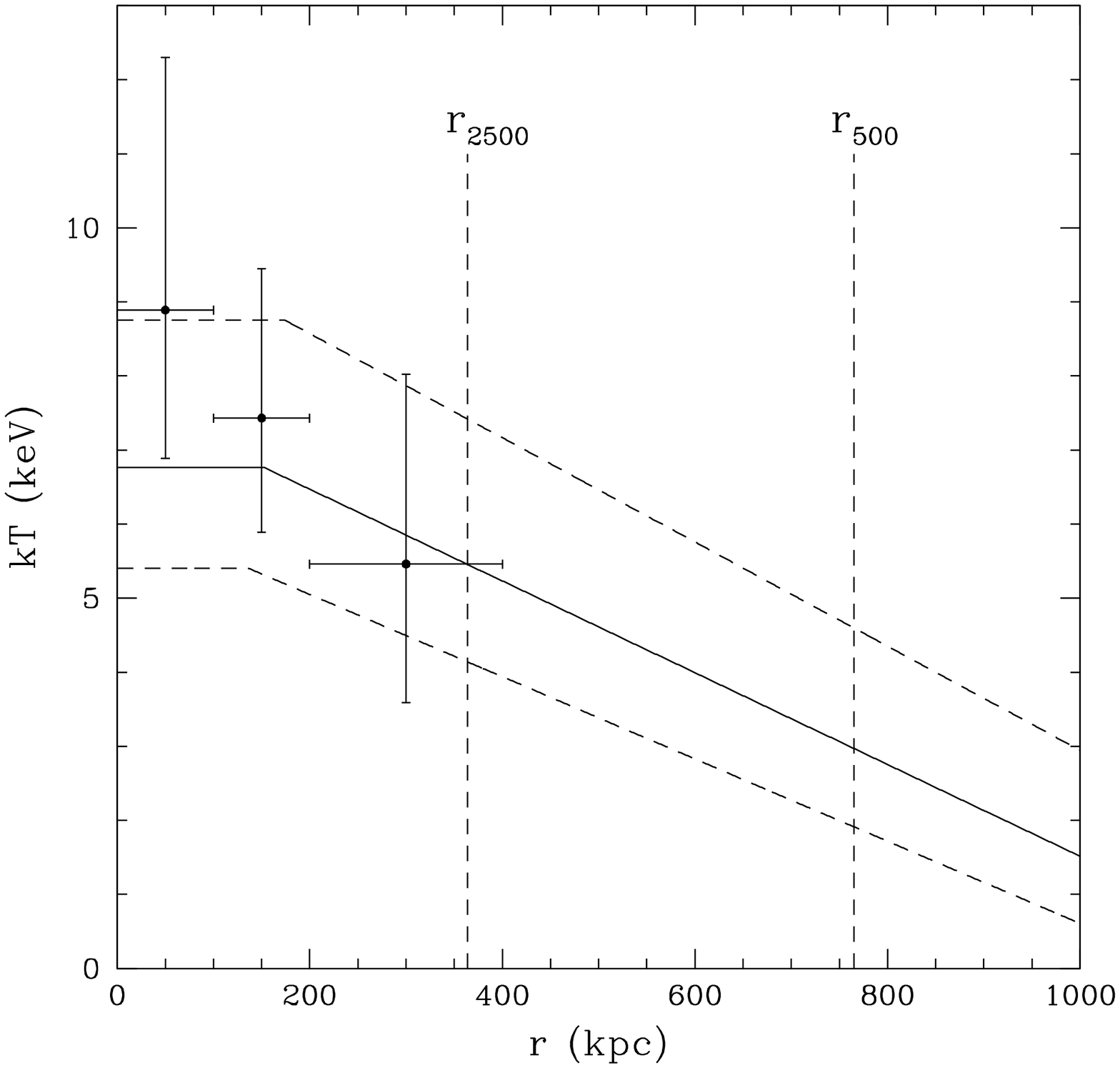}}
\caption{Temperature Profile of RCS 2318+0034 including the 1$\sigma$ error bars.
The solid line shows the broken power-law temperature profile adopted from
Vikhlinin et al. (2005) normalized by $T_{spec}$.  The dashed lines show the
broken broken power-law temperature profile normalized by the $1\sigma$
limits on $T_{spec}$.  The vertical dashed lines show $r_{2500}$ and $r_{500}$.}
\end{inlinefigure}

\noindent
background at approximately
350~kpc, it is difficult to constrain the gas temperature at larger
radii.

\section{Gas Mass, Total Mass and Gas Mass Fraction}

Deprojecting the $\beta$ model surface brightness profile, assuming
a temperature-independent emission-measure, produces a gas density profile of:

\begin{equation}
\rho(r) = \rho_0 \left(1+\left( {{r} \over {r_c}} \right)^2 \right)^{-3 \beta/2}
\end{equation}

\noindent
where $\rho_0$ is the central gas density.  Thermal plasma models
in XSPEC are normalized by the emission-measure, defined as:

\begin{equation}
\epsilon = {{10^{-14}} \over {4 \pi \left[D_A(1+z)\right]^2}}\int n_e n_H dV
\end{equation}

\noindent
where $D_A$ is the angular distance to the source,
$n_e$ is the electron number density
and $n_H$ is the hydrogen number density. The central gas
density for a given $\beta$ model is determined by inserting the density
profile given by eq. (2) into eq. (3), integrating over a cylindrical
volume with a radius of 500~kpc and equating the result
with the observed emission-measure within this region.
Using our standard $\beta$ model ($\beta=0.70$ and $r_c=142$~kpc)
and the observed emission measure within 500~kpc, we obtain
$n_e(0)=(8.04 \pm 0.18)\times 10^{-3}$~cm$^{-3}$.
The statistical uncertainty in the emission-measure produces an
uncertainty of only 2.3\% in the central gas density,
since the density scales as the square root of the emission-measure.
If we re-fit the spectrum extracted from the central 500~kpc
with the temperature fixed at 4 or 10~keV, we obtain
best-fit emission-measures of
$(3.47 \pm 0.22) \times 10^{-4}$~cm$^{-5}$ or
$(3.17 \pm 0.15) \times 10^{-4}$~cm$^{-5}$, respectively.
Thus, any variations in the gas temperature within this region
would probably not generate more than an additional
systematic uncertainty of 2.1\% in the central gas density.

\begin{inlinefigure}
\center{\includegraphics*[width=1.00\linewidth,bb=20 145 569 695,clip]{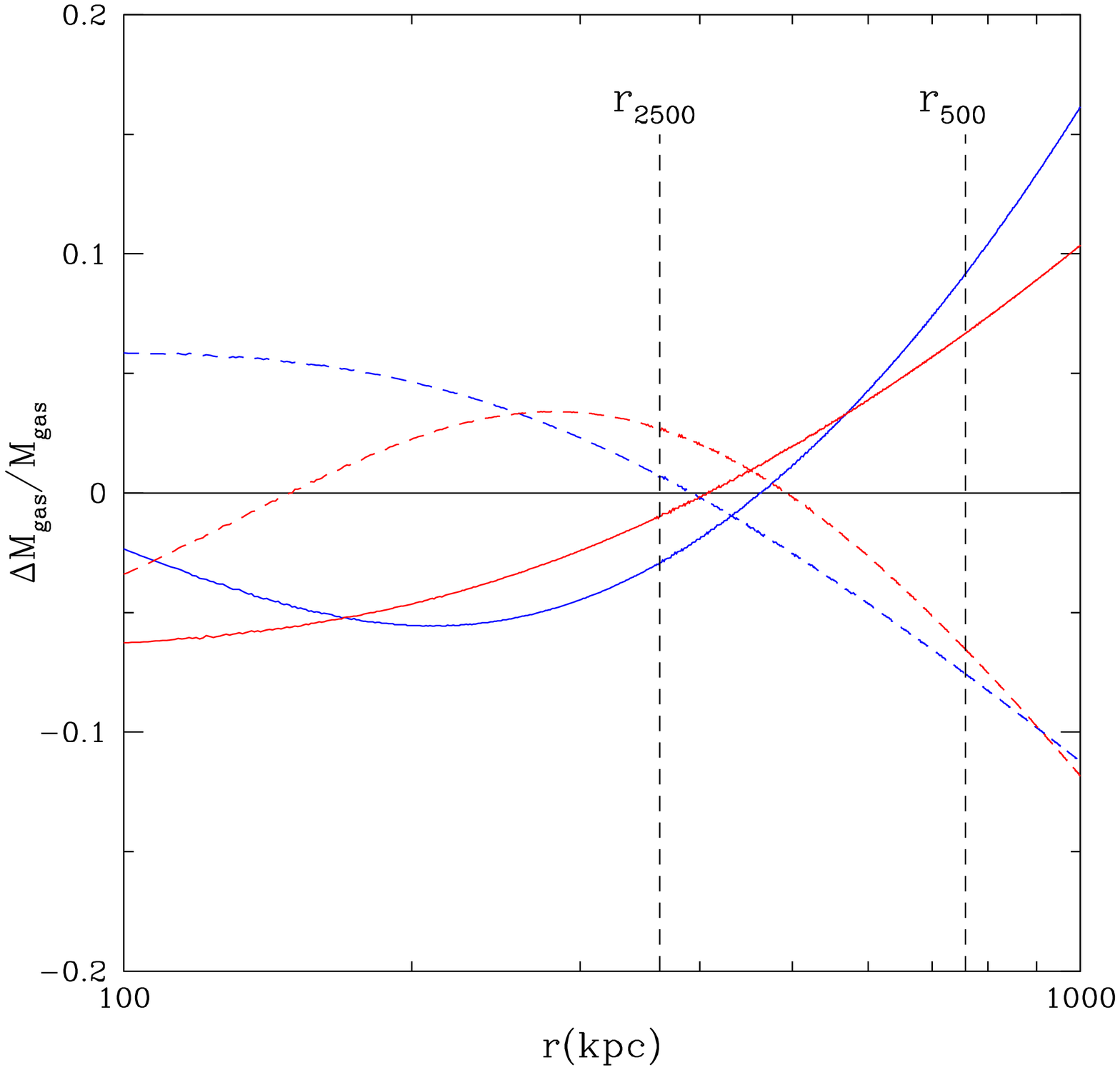}}
\caption{Ratio of the cumulative gas mass mass for several different $\beta$ models
to the cumulative gas mass derived from our standard $\beta$ model
($\beta=0.70$ and $r_c=142$~kpc).
The lines correspond to: solid red line ($\beta=0.64$ and $r_c$=136~kpc),
dashed red line ($\beta=0.85$ and $r_c$=183~kpc), sold blue line
($\beta=0.58$ and $r_c$=112~kpc) and dashed  blue line ($\beta=0.75$
and $r_c$=152~kpc). The vertical dashed lines show $r_{2500}$ and $r_{500}$.}
\end{inlinefigure}

The dominant uncertainty in the gas mass in RCS 2318+0034 is
not due to uncertainties in the central gas density, but is primarily due to
the variations in the $\beta$ model parameters derived from fitting the X-ray
surface brightness over different regions
(see the range in values of $\beta$ and $r_c$ in Tables 1 and 2).
Figure 5 shows the cumulative gas mass derived from several different
$\beta$ models relative to the cumulative gas mass derived from our
standard $\beta$ model.  The $\beta$ models used for this
comparison are given in the figure caption and are chosen to sample
the full range of possible gas masses.  Figure 5 shows that the
range of acceptable $\beta$ models produces an uncertainty in the
gas masses at $r_{2500}$ and $r_{500}$ of about 5\% and 7-10\%, respectively.
We thus assign an average combined uncertainty in the gas mass at these
radii of 10\%.

The total gravitating mass of a spherically symmetric system in
hydrostatic equilibrium is given by:

\begin{equation}
M_{tot}(<r) = {{-kTr} \over {\mu m_p G}} \left( {{d ln \rho} \over {d ln r}} + {{d ln T} \over {d ln r}} \right)
\end{equation}

\noindent
where $k$ is Boltzman's constant, $\mu=0.6$ is the mean mass per particle,
$m_p$ is the proton mass and $G$ is the gravitational constant.
The primary uncertainty in computing the gravitating mass in
RCS 2318+0034 is the lack of a detailed knowledge of the
gas temperature profile.  We therefore compute the gravitating
mass assuming two temperature profiles: 1) an isothermal profile
and 2) the broken power-law temperature profile in Vikhlinin et al. (2005).
We use our standard $\beta$ model for the gas density distribution in both cases.

Extracting a spectrum in an annulus from 70kpc to 1Mpc in RCS 2318+0034
and fitting the spectrum to an absorbed {\it apec} model
yields a temperature of $T_{spec}$=6.32 (5.05-8.18) keV ($1\sigma$ uncertainty).
The resulting gas mass, total gravitating mass and gas mass
fraction at $r_{2500}$ and $r_{500}$ assuming the gas is
isothermal are shown in Table 4.  The values for $r_{2500}$ and
$r_{500}$ 

\begin{table*}[t]
\begin{center}
\caption{Results Assuming the Gas is Isothermal}
\begin{tabular}{lcccccccc}
\hline
$\rm{T_{spec}}$ & $\rm{r_{2500}}$ & $\rm{r_{500}}$ & $\rm{M_{gas}(r_{2500})}$ & $\rm{M_{gas}(r_{500})}$ & $\rm{M_{tot}(r_{2500})}$ & $\rm{M_{tot}(r_{500})}$ & $\rm{f_{gas}(r_{2500})}$ & $\rm{f_{gas}(r_{500})}$ \\
(keV) & (Mpc) & (Mpc) & ($\rm{10^{13} M_{\odot}}$) & ($\rm{10^{13} M_{\odot}}$) & ($\rm{10^{13} M_{\odot}}$) & ($\rm{10^{13} M_{\odot}}$) & & \\
\hline\hline
5.05 & 0.308 & 0.745 & 0.796 & 2.82 & 9.96 & 28.2 & 0.080 & 0.10 \\
6.32 & 0.352 & 0.837 & 0.992 & 3.25 & 14.8 & 39.9 & 0.067 & 0.081 \\
8.18 & 0.408 & 0.955 & 1.25  & 3.81 & 23.1 & 59.3 & 0.054 & 0.064 \\
\hline
\end{tabular}
\end{center}
\noindent
Notes: The gas mass ($M_{gas}$), total gravitating mass ($M_{tot}$) and
gas mass fraction ($f_{gas}$) within $r_{2500}$ and $r_{500}$ derived from
our standard $\beta$ model and the assumption that the gas is isothermal.
Masses are shown for the best-fit emission-weighted temperature
between 70kpc and 1Mpc ($T_{spec}$) and the $1\sigma$ limits on $T_{spec}$.
\end{table*}

\begin{table*}[t]
\begin{center}
\caption{Results Using a Broken Power-Law Temperate Profile}
\begin{tabular}{lcccccccc}
\hline
$\rm{T_{spec}}$ & $\rm{r_{2500}}$ & $\rm{r_{500}}$ & $\rm{M_{gas}(r_{2500})}$ & $\rm{M_{gas}(r_{500})}$ & $\rm{M_{tot}(r_{2500})}$ & $\rm{M_{tot}(r_{500})}$ & $\rm{f_{gas}(r_{2500})}$ & $\rm{f_{gas}(r_{500})}$ \\
(keV) & (Mpc) & (Mpc) & ($\rm{10^{13} M_{\odot}}$) & ($\rm{10^{13} M_{\odot}}$) & ($\rm{10^{13} M_{\odot}}$) & ($\rm{10^{13} M_{\odot}}$) & & \\
\hline\hline
5.05 & 0.320 & 0.682 & 0.853& 2.53 & 11.2 & 21.6 & 0.076 & 0.12 \\
6.32 & 0.364 & 0.765 & 1.05 & 2.93 & 16.4 & 30.4 & 0.064 & 0.096 \\
8.18 & 0.420 & 0.872 & 1.30 & 3.43 & 25.1 & 45.1 & 0.052 & 0.076 \\
\hline
\end{tabular}
\end{center}
\noindent
Notes: The gas mass ($M_{gas}$), total gravitating mass ($M_{tot}$) and
gas mass fraction ($f_{gas}$) within $r_{2500}$ and $r_{500}$ derived from
our standard $\beta$ model along with the temperature profile in
Vikhlinin et al. (2005). Masses are shown for the the best-fit
emission-weighted temperature ($T_{spec}$) and the $1\sigma$ limits
on $T_{spec}$.
\end{table*}

\noindent
are calculated from the total gravitating mass profile.
Since the resulting
values for $r_{500}$ in Table 4 are slightly less than 1Mpc,
we also extracted spectra from several annuli with inner radii
of 70kpc and outer radii of 700, 800 and 900kpc.
Due to the limited photon statistics in the {\it Chandra} observation,
the best-fit temperatures are all consistent within the
uncertainties.  Table 4 shows that assuming an isothermal $\beta$ model
for RCS 2318+0034 produces gas mass fractions of
$0.07 \pm .02$ and $0.08 \pm .02$ at $r_{2500}$ and $r_{500}$, respectively.
The error bars take into account the 10\% uncertainty in the
gas mass noted above and the uncertainty in $T_{spec}$.
The gas mass fraction is essentially independent of radius
at large radii since, for an isothermal $\beta=0.7$ model,
$\rho_{gas} \sim r^{-2.1}$ and $\rho_{tot} \sim r^{-2}$.

Based on the analysis of {\it Chandra} observations of a sample
of 13 rich clusters of galaxies, Vikhlinin et al. (2005) found that
the temperature profile of these clusters could be fit with a
broken power-law model, given by:

\begin{equation}
T/\langle T\rangle = 
\cases{ 
1.07,&$0.035 < r/r_{180} < 0.125$,\cr
1.22 - 1.2r/r_{180},&$0.125 < r/r_{180} < 0.6$. \cr}
\end{equation}

\noindent
where $\langle T\rangle$ is the emission-weighted temperature
excluding the central 70kpc (essentially $T_{spec}$).
For clusters, $r_{500} \sim 0.6r_{180}$.  With this substitution,
the temperature profile in eq. (5) is very similar to that
derived for a sample of groups by Sun et al. (2009).

Gas masses, total gravitating masses and gas mass fractions
within $r_{2500}$ and $r_{500}$ are shown in Table 5 using
our standard $\beta$ model and the broken power-law temperature
profile in eq. (5) normalized by $T_{spec}$.
Since eq. (5) depends on $r_{500}$,
we iterated in $r_{500}$ until a self-consistent solution
was found.  The self-consistent temperature profiles are shown
in Figure 4. Table 5 shows that assuming a broken power-law
temperature profile for RCS 2318+0034 produces gas mass fractions of
$0.06 \pm .02$ and $0.10 \pm .02$ at $r_{2500}$ and $r_{500}$, respectively.
The error bars take into account the 10\% uncertainty in the
gas mass and the uncertainty in $T_{spec}$.
As opposed to the isothermal model, the gas mass fraction
increases with radius due to the declining gas temperature.

While RCS 2318+0034 is presently undergoing a merger,
Nagai et al. (2007) have shown by performing mock {\it Chandra}
observations of a sample of simulated clusters that
X-ray derived gas masses in both relaxed and unrelaxed clusters
are accurate to within 6\%.  Including this additional uncertainty in
our derived gas mass would not a have a significant affect on our estimate
of the gas mass fraction. Nagai et al. (2007) also found that
X-ray derived gravitating masses are biased low by about 5-20\%
with no trend in cluster mass or redshift. This suggests
that all X-ray derived gas mass fraction estimates are biased
high by this amount.

\section{Comparison with Previous Analysis}

Our estimated gas mass fractions using the combined 200~ksec of
{\it Chandra} data are approximately 50\% greater at $r_{2500}$ and
significantly greater at $r_{500}$ compared with the values
derived from the earlier 50~ksec observation (Hicks et al. 2008).
Comparing the derived parameters for RCS 2318+0034 in Hicks et al. with
our results, we find that both studies produced similar values
for the gas temperature ($T_{spec}=5.05-8.18$~keV compared
to $T_{spec}=5.20-7.40$~keV) and core radius ($r_c=142 \pm 23$~kpc compared
to $r_c=171 \pm 6$~kpc) for the surface brightness
profile. However, we obtain a value for the outer slope of
the surface brightness profile corresponding to $\beta=0.70 \pm .09$
compared to $\beta=0.86 \pm .03$ in Hicks et al.
This difference is primarily due to the poorer
photon statistics in the initial 50~ksec {\it Chandra} observation.
Larger values of $\beta$ will over estimate the total gravitating mass
and underestimate the gas mass and gas mass fraction.
Jones \& Forman (1999) found an average value of $\beta=0.6$
when fitting the surface brightness profile of a large
sample of rich clusters observed with 

\begin{inlinefigure}
\center{\includegraphics*[width=1.00\linewidth,bb=20 145 569 695,clip]{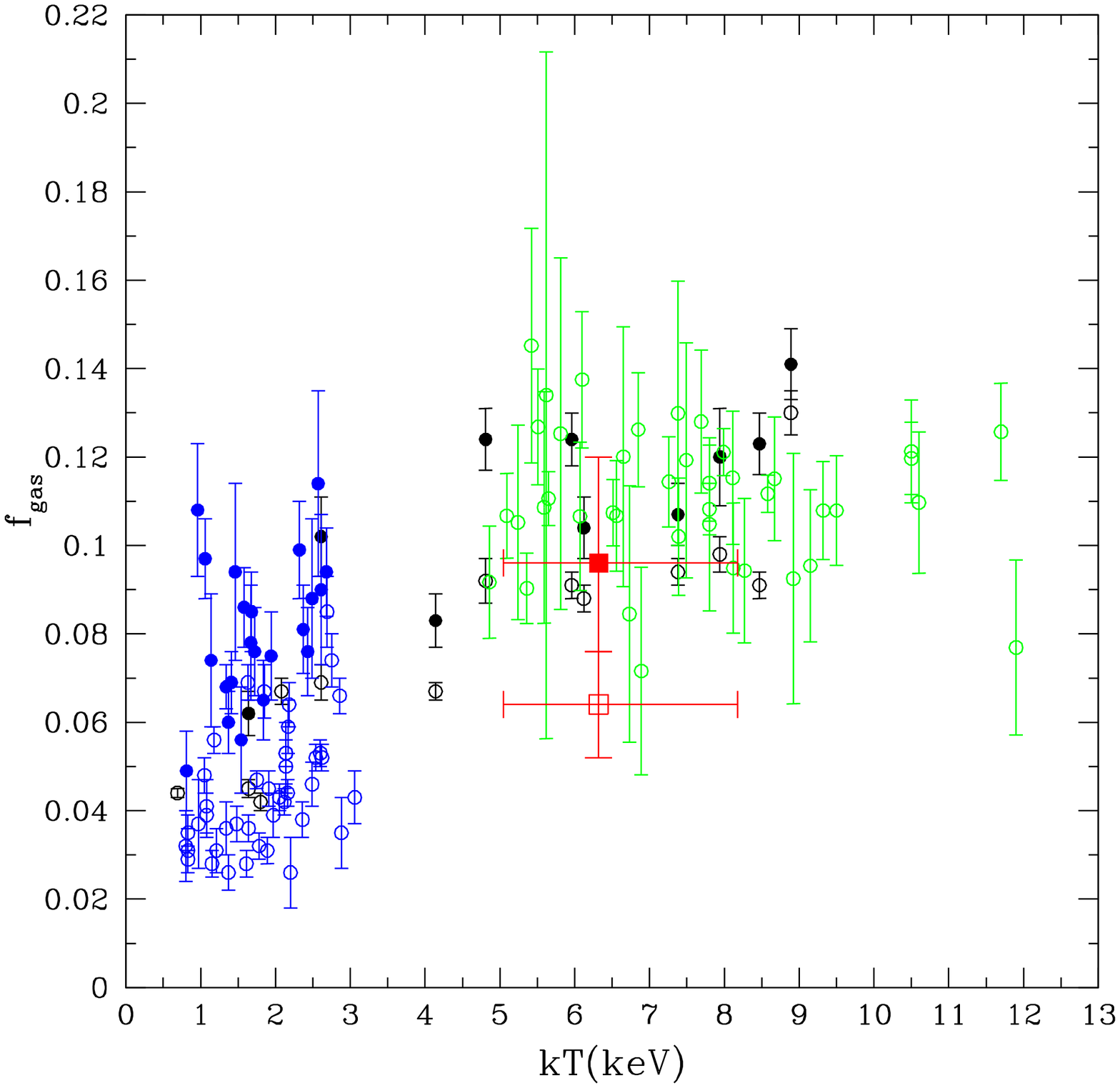}}
\caption{Gas mass fractions from Sun et al. (2009) (blue), Vikhlinin et al. (2006)
(black), Allen et al. (2009) (green) and RCS 2318+0034 (red).  Open symbols
give $f_{gas}$ within $r_{2500}$ and solid symbols give $f_{gas}$ within $r_{500}$.}
\end{inlinefigure}

\noindent
the {\it Einstein}
Observatory. More recently, in a study of 31 clusters culled
from the REXCESS sample, which includes both relaxed and
unrelaxed clusters, Croston et al. (2008) found an average
power-law slope in the surface
brightness profile between $0.3r_{500}$ and $0.8r_{500}$
among 6~keV clusters corresponding to $\beta=0.7$.

At $r_{2500}$, we obtain
$M_{gas}(r_{2500}) = 9.9^{+2.6}_{-2.0} \times 10^{12} \Mo$ and
$M_{tot}(r_{2500})= 1.5^{+0.8}_{-0.5} \times 10^{14}~\Mo$
assuming the gas is isothermal.  Hicks et al. (2008) derived
$M_{gas}(r_{2500}) = (9.7 \pm 0.6) \times 10^{12} \Mo$
and $M_{tot}(r_{2500})= (2.43 \pm 0.28) \times 10^{14}~\Mo$
assuming the gas is isothermal.  However, the gas and
total masses listed for RCS 2318+0034 in Table 7
of Hicks et al. are inconsistent with the parameters given
in Tables 3 and 4.  The corrected values are
$M_{gas}(r_{2500}) = 1.60 \times 10^{13} \Mo$
and $M_{tot}(r_{2500})= 2.80 \times 10^{14}~\Mo$ (Hicks 2011, private
communication).  While there are some differences between our results due
to differences in $r_{2500}$ and $\beta$, our gas mass
fraction at $r_{2500}$ is consistent with the corrected Hicks (2011,
private communication) result.
We find that assuming a broken power-law temperature profile increases
both $M_{gas}(r_{2500})$ and $M_{tot}(r_{2500})$ by 10\% relative to the values
derived by assuming the gas is isothermal.
Thus, both temperature profiles produce the same value
for the gas mass fraction of $f_{gas}(r_{2500})=0.06 \pm 0.02$.

At $r_{500}$, we obtain
$M_{gas}(r_{500})= 3.2^{+1.0}_{-0.7}  \times 10^{13} \Mo$,
$M_{tot}(r_{500})= 4.0^{+1.9}_{-1.2}  \times 10^{14}~\Mo$ and
$f_{gas}(r_{500}=0.06 \pm 0.02$
assuming the gas is isothermal.  Hicks et al. (2008)  derived
$M_{gas}(r_{500}) = (1.99 \pm 0.28) \times 10^{13} \Mo$
and $M_{tot}(r_{500})= (12.9 \pm 2.0) \times 10^{14}~\Mo$.
However, the corrected values are
$M_{gas}(r_{500}) = 3.64 \times 10^{13} \Mo$
and $M_{tot}(r_{500})= 6.61 \times 10^{14}~\Mo$ (Hicks 2011, private
communication).
Assuming a broken power-law temperature profile reduces $M_{gas}(r_{500})$
by 10\% (due to the smaller value of $r_{500}$) and
$M_{tot}(r_{500})$ by 25\% (primarily due to the declining gas temperature).
The assumption of a broken power-law temperature profile
thus produces a larger gas mass fraction of $f_{gas}(r_{500})=0.10 \pm 0.02$.

\begin{inlinefigure}
\center{\includegraphics*[width=1.00\linewidth,bb=20 145 569 695,clip]{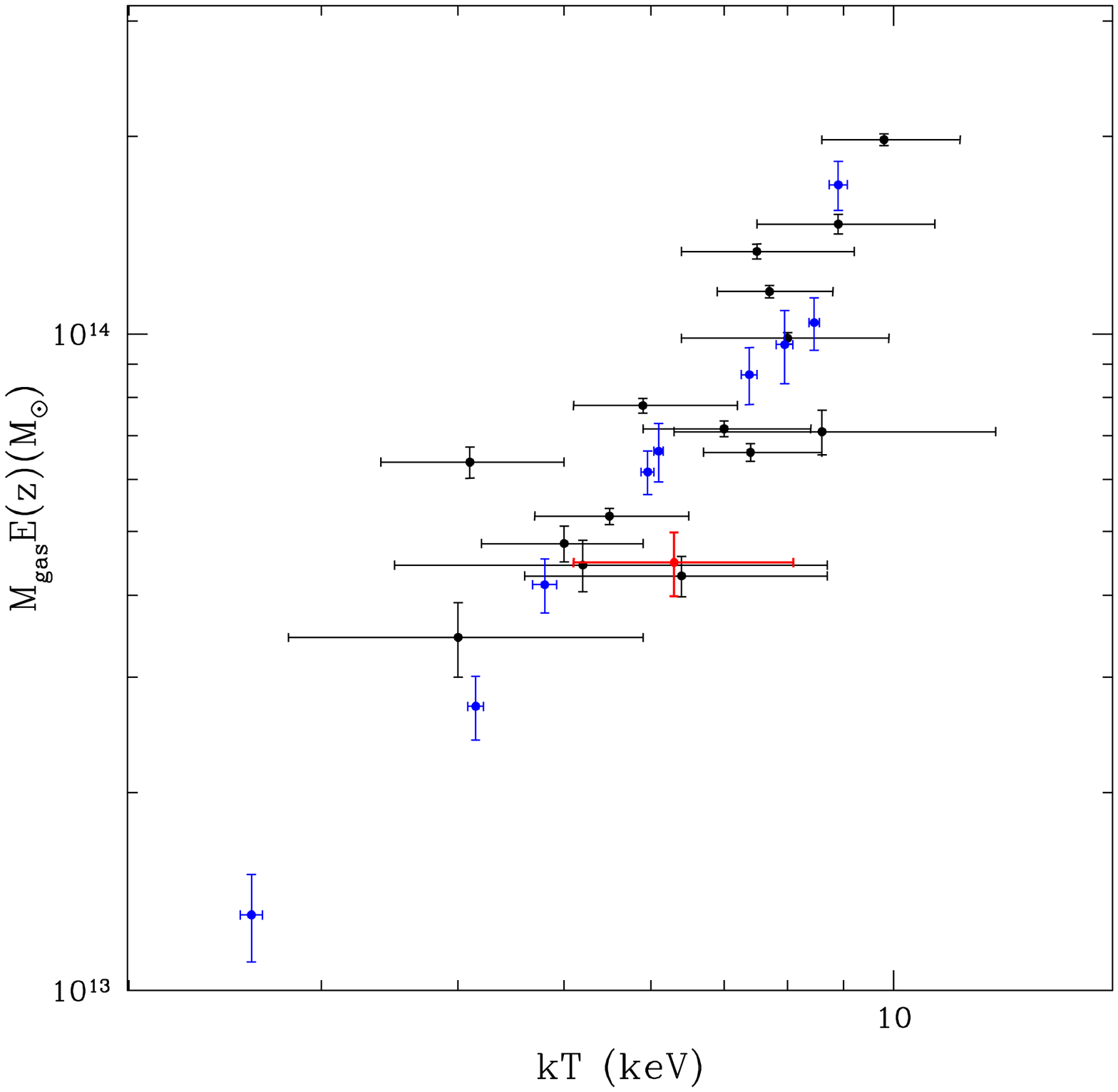}}
\caption{Scaled gas mass vs. spectroscopic temperature for the
X-ray selected clusters in Vikhlinin et al. (2006) (blue data points), the SZ-selected
clusters in Andersson et al. (2010) (black data points) and the optically-selected
cluster RCS 2318+0034 (red data point).}
\end{inlinefigure}

\section{Discussion and Summary}

In general, the gas mass fraction in groups and clusters
depends on the radius at which the gas mass fraction is
computed and the virial mass of the cluster. At small radii,
the gas mass fraction also depends on whether the cluster has
a cool-core.  The central gas density in cool-core clusters is more
than an order of magnitude greater than the central density in
non-cool core systems.  Since $r_{2500}$ is only $\sim$~200~kpc
in groups, the presence of a cool-core can have a significant
impact on the gas mass fraction at $r_{2500}$ in groups.
Sun et al. (2206) shows that there is indeed a strong anti-correlation
between the gas mass fraction in groups at $r_{2500}$ and
the central gas entropy.  For hotter systems (larger values of $r_{2500}$),
the presence of a cool-core has less of an impact on
the gas mass fraction at $r_{2500}$.

Reports of the average gas mass fraction within $r_{2500}$ vary from
4.5\% in groups (Sun et al. 2009), up to 7.8\% (Vikhlinin et al. 2006)
and 11.0\% (Allen et al. 2008) in clusters.
The reported gas mass fraction within $r_{500}$
varies from 8.1\% in groups (Sun et al. 2009), up to
13.0\% in clusters (Vikhlinin et al. 2006, Ettori et al. 2010,
Pratt et al. 2010, Juett et al. 2010).
These results are illustrated in Figure 6 and
show that the differences are mainly due to the range
in gas temperatures, or cluster masses, sampled in these
different studies.  The average gas temperature of the clusters in the
Vikhlinin et al. sample is 4.5~keV compare to
7.7~keV in the sample of clusters analyzed by Allen et al.,
which contains a substantial fraction of clusters culled from the
Massive Cluster Survey (MACS; Ebeling et al. 2001).

Our estimate for the gas mass fraction within $r_{500}$ in
the optically selected cluster, RCS2318+0034, is
fully consistent with the results of X-ray selected clusters (see Figure 6).
This result is also illustrated in Figure 7, where we compare the
scaled gas mass and spectroscopic temperature within $r_{500}$
for the X-ray selected clusters in Vikhlinin et al. (2006),
the SZ-selected clusters
in Andersson et al. (2010) and the optically selected cluster,
RCS 2318+0034.  Croston et al. (2008) derived a

\begin{inlinefigure}
\center{\includegraphics*[width=1.00\linewidth,bb=20 145 569 695,clip]{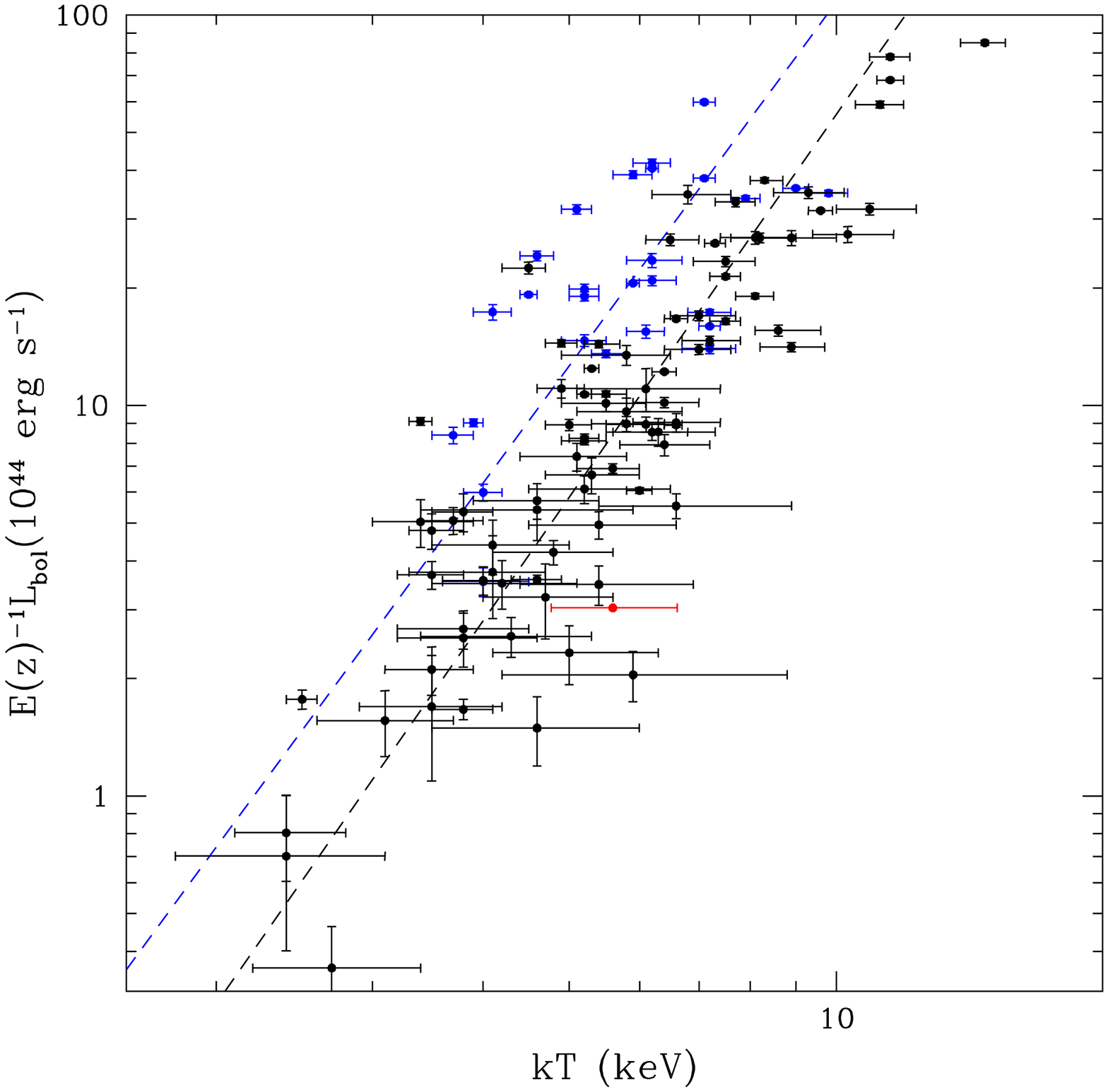}}
\caption{Bolometric X-ray luminosity vs. cluster temperature relation for the
total emission within $r_{500}$.  The RCS 2318+0034 data point is shown
in red.  All other data points are from Maughan et al. (2011).  Relaxed clusters
are shown in blue and unrelaxed clusters are shown in black.  Also shown
are the best fit relations for relaxed clusters (dashed blue line)
and unrelaxed clusters (dashed black line) derived in Maughan et al.}
\end{inlinefigure}

\noindent
scaling relation
between the gas mass and temperature based on an analysis of
clusters culled from the REXCESS sample. The scaling
relation in  Croston et al. is based on the temperature
between $0.15r_{500}$ and $0.75r_{500}$. Extracting a spectrum
from within this region in RCS2318+0034 gives a temperature
of $5.3 \pm 091$~keV.  Using the scaling relations in Croston et al. gives
an estimated gas mass at $r_{500}$ of $(2.2-4.2) \times 10^{13} \Mo$,
which is consistent with our estimate for the gas mass in RCS2318+0034
(see Table 4).  In addition, the scaling relation between $f_{gas}(r_{500})$
and $M_{tot}(r_{500})$ derived by Vikhlinin et al. (2009)
predicts $f_{gas}(r_{500})$=10.0\%, which is in very good
agreement with our result.  Within $r_{2500}$, we find a gas mass fraction of
6\% for RCS2318+0034, which is significantly less than that found for
most clusters (see Figure 6).  Figures 1 and 2 show that RCS2318+0034
is undergoing a merger which may affect the estimate
of the gas mass fraction at $r_{2500}$.

One of the most extensively studied cluster relations is that between
the gas temperature and X-ray luminosity (e.g., Edge \& Stewart 1991,
David et al. 1993; Fabian et al. 1994; Markevitch 1998,
Arnaud \& Evrard 1999). Early studies found a great deal of scatter
in the Lx-T relation (Edge \& Stewart 1991; David et al. 1993).
Fabian et al. (1994) showed that most of this scatter was due
to the inclusion of massive cooling flow clusters along with non cool-core
clusters in the previous studies. The X-ray luminosity of massive cooling
flow clusters can be up to five times greater than the X-ray luminosity
of non cool-core clusters for a given temperature.
By removing the emission from cool cores, Markevitch (1998) found that
the scatter in the Lx-T relation can be greatly reduced.  Similar conclusions
have also been obtained by more recent studies
(e.g., Maughan 2007, Prat et al. 2009, Maughan et al. 2011).
While most studies have shown that the Lx-T relation for
clusters at a given redshift is not self-similar, there is
significant evidence that the evolution of the Lx-T relation
is nearly self-similar (e.g., Vikhlinin et al. 2002, Maughan 2006,
Maughan et al. 2011).

Maughan et al. (2011) recently studied the Lx-T relation for a
sample of 114 clusters observed by {\it Chandra} with redshifts
of $0.1 < z < 1.3$.  The Lx-T relations in Maughan et al. were derived
from both the total emission within $r_{500}$ and the emission
within $(0.15-1.0)r_{500}$.  Due to the limited
photon statistics in the {\it Chandra} observation of
RCS2318+0034, the uncertainty in the gas temperature
between $(0.15-1.0)r_{500}$ is too large for a useful
comparison, so we only compare results using the total
emission within $r_{500}$.  Figure 8 shows the scaled Lx-T data
for the relaxed and unrelaxed clusters in Maughan et al. (2011)
along with the temperature and X-ray luminosity of RCS2318+0034.
Figure 8 shows that the temperature and X-ray luminosity
of the z=0.78 cluster, RCS2318+0034, is consistent with
other unrelaxed clusters with similar redshifts
given the observed scatter among clusters.

In summary, we find that the large scale properties of the
optically selected cluster RCS2318+0034 are consistent with
the large scale properties of X-ray selected clusters.
At smaller radii, we find a lower gas mass fraction
in RCS2318+0034 compared to most X-ray selected clusters,
which may be due to the on-going merger.

We would like to than A. Hicks for her help with comparing
our results and also A. Vikhlinin for useful comments.  This work
was supported in part by the NASA grant GO8-9122X.


\begin{references}
\reference{}Allen, S., Schmidt, R., Ebeling, H., Fabian, A., van Speybroeck, L. 2004, MNRAS, 353, 457.
\reference{}Allen, S., Rapetti, D., Schmidt, R., Ebeling, H., Morris, R. \& Fabian, A. 2008, MNRAS, 383, 879.
\reference{}Andersson, K. et al. 2010 (astro-ph 1006.3068).
\reference{}Arnaud, M. \& Evrard, A. 1999, MNRAS, 305, 631.
\reference{}Ascasibar, Y. \& Markevitch, M. 2006, ApJ, 650, 102.
\reference{}Bialek, J., Evrard, A. E., Mohr, J. 2001, ApJ, 555, 597.
\reference{}Bode, P., Ostriker, J. \& Vikhlinin, A. 2009, ApJ, 700, 989.
\reference{}Croston, J., Pratt, G., Bohringer, H., Arnaud, M., Pointecouteau, E., Ponman, T., Sanderson, A., Temple, R., Bower, R. \& Donahue, M. 2008, A\&A, 487, 431.
\reference{}Dav\'{e}, R., Oppenheimer, B., Sivanandam \& S. 2008, MNRAS, 391 110.
\reference{}David, L., Arnaud, K., Forman, W. \& Jones, C. 1990, ApJ, 356, 32.
\reference{}David, L., Slyz, A., Jones, C., Forman, W., Vrtilek, S. \&  Arnaud, K. 1993, ApJ, 412, 479. 
\reference{}David, L., Jones, C. \& Forman, W. 1995, ApJ, 445, 578.
\reference{}Ebeling, H., Edge, A. \& Henry, J. 2001, ApJ, 553, 668.
\reference{}Edge, A. \& Stewart, G. 1991, MNRAS, 252, 414.
\reference{}Ettori, S., Tozzi, P. \& Rosati, P. 2003, A\&A, 398, 879.
\reference{}Ettori, S., Morandi, A., Tozzi, P., Balestra, I., Borgani, S., Rosati, P., Lovisari, L.\& Terenziani, F. 2009, A\&A, 501, 61.
\reference{}Ettori, S., Gastaldello, F., Leccardi, A., Molendi, S., Rossetti, M., Buote, D. \& Meneghetti, M. 2010, A\&A, 524, 68.
\reference{}Fabian, A., Crawford, C., Edge, A. \& Mushotzky, R. 2004, MNRAS, 267, 779.
\reference{}Ferramacho, L. \& Blanchard, A. 2007, A\&A, 463, 423.
\reference{}Finoguenov, A.; Jones, C., B\"{o}hringer, H. \& Ponman, T. 2002, ApJ, 578, 74.
\reference{}Gladders, M. \& Yee, H. 2005, ApJS, 157, 1.
\reference{}Gonzalez, A., Zaritsky, D., Zabludoff, A. 2007, ApJ, 666, 147.
\reference{}Hicks, A., Ellingson, E., Bautz, M., Cain, B., Gilbank, D., Gladders, M., Hoekstra, H., Yee, H. \& Garmire, G. 2008, ApJ, 680, 1022.
\reference{}Hicks, A., 2011, private communication.
\reference{}Jones, C. \& Forman, W. 1999, ApJ., 511, 65.
\reference{}Juett, A., Davis, D. \& Mushotzky, R. 2010, ApJ, 709, 103.
\reference{}Kravtsov, A., Andrey V., Nagai, D. \& Vikhlinin, A. 2005, 2005, ApJ, 625, 588.
\reference{}Lloyd-Davies, E., Ponman, T. \& Cannon, D. 2000, MNRAS, 315, 689.
\reference{}Markevitch, M. 1998, ApJ, 504, 27.
\reference{}Maughan, B. 2007, ApJ, 668, 772.
\reference{}Maughan, B., Giles, P., Randall, S., Jones, C. \& Forman, W. 2011 (astro-ph 1108.1200).
\reference{}Nagai D., Vikhlinin, A. \& Kravtsov, A. 2007, ApJ, 655, 98.
\reference{}Neumann, D. \& Arnaud, M. 2001, A\&A, 373, 33.
\reference{}Ponman, T., Cannon, D. \& Navarro, J. 1999, Nature, 397, 135.
\reference{}Pratt, G., Croston, Arnaud, M. \& Bohringer, H. 2009, A\&A, 361, 378.
\reference{}Pratt, G., Arnaud, M., Piffaretti, R., Bohringer, H., Ponman, T., Croston, J., Voit, G., Borgani, S. \& Bower, R. 2010, A\&A, 511, 85. 
\reference{}Sun, M., Voit, G., Donahue, M., Jones, C., Forman, W. \& Vikhlinin, A. 2009, ApJ, 693, 1142.
\reference{}Vikhlinin, A., van Speybroeck, L., Markevitch, M., Forman, W. \& Grego, L. 2002, ApJ, 578, 107.
\reference{}Vikhlinin, A., Kravtsov, A., Forman, W., Jones, C., Markevitch, M., Murray, S. \& Van Speybroeck, L. 2006, ApJ, 640, 691. 
\reference{}Vikhlinin, A., Burenin, R., Ebeling, H., Forman, W., Hornstrup, A., Jones, C., Kravtsov, A., Murray, S., Nagai, D., Quintana, H. \& Voevodkin, A. 2009, ApJ, 692, 1033.
\reference{}Voit, G. M., Kay, S. \& Bryan, G. 2005, MNRAS, 364, 909.
\reference{}White, D. \& Fabian, A. 1995, MNRAS, 273, 72.
\reference{}White, S. Navarro, J., Evrard, A. E. \& Frenk, C. 1993, Nature, 366, 429.
\end{references}
\end{document}